\documentclass[aps,prl,twocolumn,groupedaddress,showpacs,showkeys,superscriptaddress,nofootinbib]{revtex4}%
\usepackage{graphics}%
\begin{document}%
\def\Barcelo{Barcel\'o}
\title{Hawking-like radiation does not require a trapped region}
\author{C. \Barcelo}
\affiliation{Instituto de Astrof\'{\i}sica de Andaluc\'{\i}a, CSIC,
Camino Bajo de Hu\'etor 50, 18008 Granada, Spain}
\author{S. Liberati}
\affiliation{International School for Advanced Studies, Via Beirut
2-4, 34014 Trieste, Italy and INFN, Sezione di Trieste}
\author{S. Sonego}
\affiliation{Universit\`a di Udine, Via delle Scienze 208, 33100 Udine, Italy}
\author{M. Visser}
\affiliation{School of Mathematics, Statistics, and Computer Science,
Victoria University of Wellington, New Zealand}
\date{22 June 2006; 
\LaTeX-ed \today}
\bigskip
\begin{abstract}

We discuss the issue of quasi-particle production by ``analogue
black holes'' with particular attention to the possibility of
reproducing Hawking radiation in a laboratory.  By constructing
simple geometric acoustic models, we obtain a somewhat unexpected
result: We show that in order to obtain a stationary and Planckian
emission of quasi-particles, it is \emph{not} necessary to create a
trapped region in the acoustic spacetime (corresponding to a
supersonic regime in the fluid flow).  It is sufficient to set up a
dynamically changing flow asymptotically approaching a sonic regime
with sufficient rapidity in laboratory time.

\end{abstract}
\maketitle

\def\e{{\mathrm e}}%
\def\g{{\mbox{\sl g}}}%
\def\Box{\nabla^2}%
\def\d{{\mathrm d}}%
\def\R{{\rm I\!R}}%
\def\ie{{\em i.e.\/}}%
\def\eg{{\em e.g.\/}}%
\def\etc{{\em etc.\/}}%
\def\etal{{\em et al.\/}}%
\def\HRULE{{\bigskip\hrule\bigskip}}
\noindent{\underline{\em Introduction}:} It is by now well
established that the physics associated with classical and quantum
fields in curved spacetimes can be reproduced, within certain
approximations, in a variety of different physical systems by
constructing so-called ``analogue
spacetimes''~\cite{analogue-book,living-review}. Among such systems
the simplest example is provided by a barotropic, irrotational and
viscosity-free fluid. Small classical (or quantum) disturbances
propagating in this type of fluid behave equivalently to a linear
classical (or quantum) field over an effective {\em acoustic
spacetime\/}, endowed with an {\em acoustic metric\/}. In the case
of a fluid flow along an infinitely long thin pipe, with density and
velocity fields constant on any cross section orthogonal to the pipe,
one can write the $(1+1)$-dimensional acoustic metric as%
\begin{equation}%
\g=\Omega^2\left[-\left(c^2-v^2\right)\d t^2
-2\,v\,\d t\,\d x+\d x^2\right]
\;,%
\label{metric}%
\end{equation}%
where $t\in\R$ denotes laboratory time and $x\in\R$ the physical
distance along the pipe. In this expression $c$ is the speed of
sound, $v$ is the fluid velocity, and $\Omega$ is an unspecified
non-vanishing function~\cite{visser98}.  In general, all these
quantities depend on the laboratory coordinates $x$ and $t$. Here,
we shall assume that $c$ is a constant. Hence, it is the velocity
$v(x,t)$ that contains all the relevant information about the causal
structure of the acoustic spacetime.  (See Ref.~\cite{companion} for
a detailed analysis of the causal structure associated with a broad
class of $(1+1)$-dimensional acoustic geometries, both static
and dynamic.)%

In this letter we analyze, in simple terms, the issue of quantum
quasi-particle creation (in this specific case, phonon creation) by
externally specified $(1+1)$-dimensional analogue geometries
simulating the formation of black hole-like configurations. (Several
additional cases and a more extensive discussion of the results can
be found in Ref.~\cite{quasi-particle-cqg}.)  In this analysis we
have in mind, on the one hand, the possibility of setting up
laboratory experiments exhibiting Hawking-like
radiation~\cite{hawking,bd} and, on the other hand, the acquisition
of new insights into the physics of black hole evaporation in
semiclassical gravity.  In particular, we have found, quite
surprisingly, that in order to produce a Hawking-like effect it is
neither necessary to generate a supersonic regime (fluid velocity
$v$ strictly larger than sound velocity $c$), nor even a sonic point
at finite laboratory time. All one needs is that a sonic point
develops in the asymptotic future (that is, for $t\to +\infty$)
\emph{with sufficient rapidity} (we shall in due course explain
exactly what we mean by this).%

\noindent{\underline{\em General framework}:} A sonic point in the
flow, where $v(t,x)=\pm c$, corresponds to a so-called acoustic
trapping (or apparent) horizon for the Lorentzian geometry defined
by the metric (\ref{metric}). See \eg\ Ref.~\cite{living-review},
Sec.~2.5, pp.~15--16.  Consider a monotonically non-decreasing
function $\bar{v}(x)$ such that $\bar{v}(0)=-c$ and $\bar{v}(x)\to
0$ for $x\to +\infty$.  If one chooses $v(x,t)=\bar{v}(x)$ in
(\ref{metric}), the corresponding acoustic spacetime represents, for
observers with $x>0$, a static black hole with the horizon located
at $x=0$ (in this case trapping and event horizon coincide), a black
hole region for $x<0$, and a (right-sided) surface
gravity~\cite{visser98}%
\begin{equation}%
\kappa:=\lim_{x\to 0^+}\frac{\d \bar{v}(x)}{\d x}\;.%
\label{kappa}%
\end{equation}%
\begin{figure*}[t]%
\vbox{ \hfil \scalebox{0.70}[0.60]{ {\includegraphics{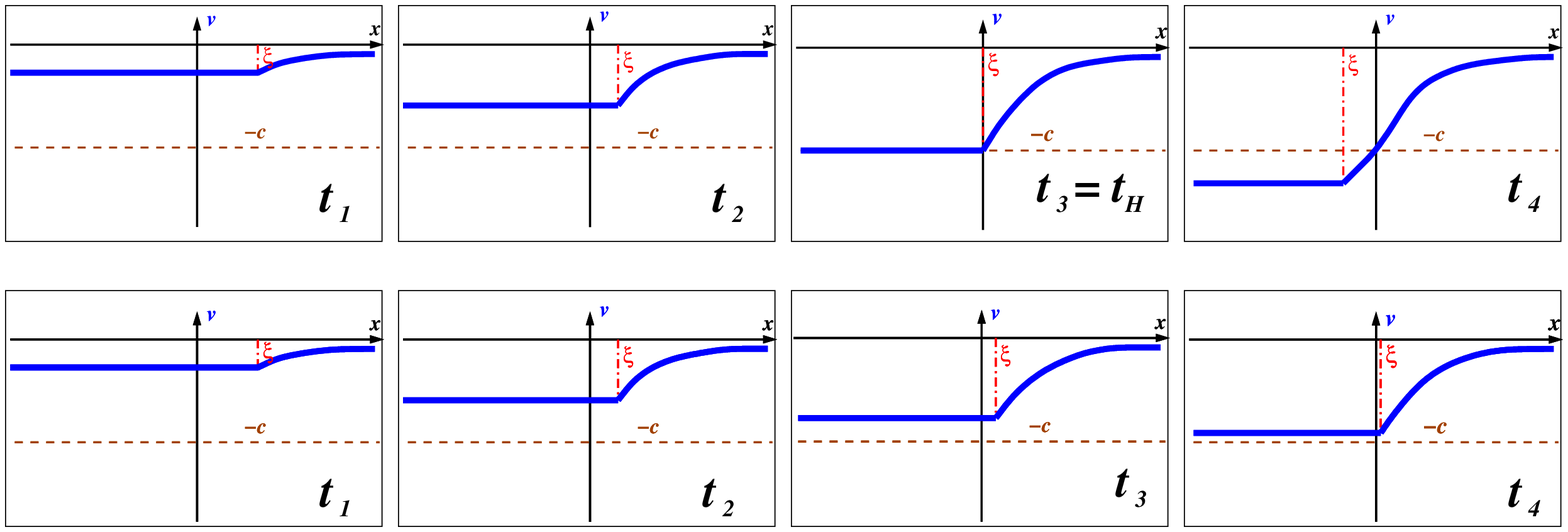}} }\hfil}%
\bigskip%
\caption{%
  The plot shows a series of temporal snapshots,
  $t_{1}<t_{2}<t_{3}<t_{4}$, of the velocity profile $v(x,t)$. In
  the first-row series, the sonic regime is reached at a finite
  laboratory time (this series represents the formation of a
  non-extremal acoustic black hole). In the second-row series, the
  sonic regime is reached only asymptotically in laboratory time (this
  series represents the formation of what we called a critical
  acoustic black hole).  }
\label{F:xi-cbh}%
\end{figure*}%

Now, taking the above $\bar{v}(x)$, let us consider $t$-dependent
velocity functions%
\begin{equation}%
v(x,t)=\left\{\begin{array}{lll}%
\bar{v}(\xi(t))&\mbox{if}& x\leq\xi(t)\;,%
\\ %
\bar{v}(x)&\mbox{if}& x\geq\xi(t)\;,%
\end{array}\right.%
\label{velocity}%
\end{equation}%
with $\xi$ a monotonically decreasing function of $t$, such that
$\displaystyle\lim_{t \to -\infty} \xi(t) = +\infty$ and
$\displaystyle\lim_{t \to -\infty} \dot \xi(t) = 0$.  (The first
condition serves to guarantee that spacetime is flat at early times,
whereas we impose the second one only for simplicity.)  The
derivative with respect to $x$ of the corresponding velocity profile
is discontinuous at $x=\xi(t)$, where a ``kink'' traveling to the
left is located (see Fig.~\ref{F:xi-cbh}).  There are basically two
possibilities for $\xi$, according to whether the value $\xi=0$ is
attained for a finite laboratory time $t_{\rm H}$ or asymptotically
for an infinite future value of laboratory time.%

In the first case $\xi(t_{\rm H})=0$ and the corresponding metric
(\ref{metric}) represents the formation of a non-extremal black hole
(in this letter we will consider only the case $\kappa \neq 0$,
extremal black holes --- zero surface gravity --- are treated
in~\cite{quasi-particle-cqg}).  For small values of $|t-t_{\rm H}|$
we have%
\begin{equation}%
\xi(t)=-\lambda\,(t-t_{\rm H})+{\cal O}([t-t_{\rm H}]^2)\;,%
\label{appxi-bh}%
\end{equation}%
where $\lambda$ is a positive parameter.  Apart from this feature, the
detailed behaviour of $\xi$ is largely irrelevant for our purposes.%

If instead $\xi\to 0$ is attained only at infinite future time, that
is $\displaystyle\lim_{t\to +\infty}\xi(t)=0$, one is describing the
asymptotic formation of what we shall call a critical black hole. In
the following we shall restrict the discussion to the case in which
for $t\to +\infty$, the behaviour of $\xi$ is exponential, that is
$\xi(t)\sim A\,\e^{-\kappa_{\rm D}t}$, with $\kappa_{\rm D}$ a
positive constant, in general different from $\kappa$, and $A>0$. Of
course, other possibilities for the asymptotics can be envisaged;
for example, in Ref.~\cite{quasi-particle-cqg} the case of a power
law $\xi(t)\sim B\,t^{-\nu}$, with $\nu>0$ and $B>0$, is also
studied in detail.%

Starting with a quantum scalar field in its natural Minkowskian
vacuum at $t\to -\infty$, we want to know the total quantity of
quasi-particle production to be detected at the right asymptotic
region at late times, $t\to +\infty$, caused by the dynamical
evolution of the velocity profile $v(x,t)$.  This can be done by
following the standard procedure in which one computes the
Bogoliubov $\beta$ coefficients between modes that have positive
frequency with respect to null coordinates $U$ and $u$, with $U$
regular on the acoustic event horizon $\cal H$, whereas $u$ tends to
$+\infty$ as $\cal H$ is approached.  (See, \eg, Ref.~\cite{bd}.) We
must thus find the relation between $U$ and $u$ for a sound ray that
is close to the horizon, \ie, in the asymptotic regime $u\to
+\infty$.  If this relation turns out to be exponential, it is then
a well established result that a Planckian spectrum will be observed
at late
times~\cite{hu}, so Hawking-like radiation is recovered.%

In the geometric acoustic approximation, a right-going sound ray is
an integral curve of the differential equation%
\begin{equation}%
\frac{\d x}{\d t}=c+v(x,t)\;,%
\label{diffeq}
\end{equation}%
which can be integrated exactly in order to find the values of $U$
and $u$ for a generic right-moving ray that crosses, at laboratory
time $t_0$, the kink $x=\xi(t)$ (details can be found in
Ref.~\cite{quasi-particle-cqg}):%
\begin{equation}%
U=t_0-\frac{\xi(t_0)}{c}+\frac{1}{c}\int_{-\infty}^{t_0}\d
t\,\bar{v}(\xi(t))\;;%
\label{UU}%
\end{equation}%
\begin{equation}%
u=t_0-\frac{\xi(t_0)}{c}-\frac{1}{c}\int_{\xi(t_0)}^{+\infty}\d
x\,\frac{\bar{v}(x)}{c+\bar{v}(x)}\;.%
\label{uu}%
\end{equation}%
Eliminating $t_0$ between Eqs.~(\ref{UU}) and (\ref{uu}) provides
the sought-for relation between $U$ and $u$.%

\noindent{\underline{\em Formation of a non-extremal black hole}:}
In the case of the formation of a non-extremal black hole, the
function $\bar{v}(x)$ must be such that for small values of $|x|$, one
can write%
\begin{equation}%
\bar{v}(x)=-c+\kappa\,x+{\cal O}(x^2)\;.%
\label{appv-bh}%
\end{equation}%
In addition, the trapping horizon forms at a finite laboratory time, say at
$t=t_{\rm H}$, just when $\xi(t_{\rm H})=0$. In this situation
an event horizon always exists, generated by the
right-moving ray that eventually remains frozen on the trapping
horizon, at $x=0$. For such a ray $t_0\to t_{\rm H}$, and since
$\xi(t_{\rm H})=0$, the $U$ parameter has the finite value%
\begin{equation}%
U_{\rm H}= t_{\rm H} +
\frac{1}{c}\int_{-\infty}^{t_{\rm H}}\d t\,\bar{v}(\xi(t))\;.%
\label{UH-fin}%
\end{equation}%
For a ray with $U<U_{\rm H}$ we then obtain, combining
Eqs.~(\ref{UU}) and (\ref{UH-fin}):%
\begin{equation}%
U=U_{\rm H}+t_0-t_{\rm H}-\frac{\xi(t_0)}{c}
-\frac{1}{c}\int_{t_0}^{t_{\rm H}}\d t\,\bar{v}(\xi(t))\;.%
\label{U-fin}%
\end{equation}%
This \emph{exact} equation is now in a form suitable for
conveniently extracting \emph{approximate} results in the region
$t_0\sim t_{\rm H}$, corresponding to sound rays that ``skim'' the
horizon. Using Eqs.~(\ref{appxi-bh}) and (\ref{appv-bh}) we find
$U=U_{\rm H}+\frac{\lambda}{c}\,(t_0-t_{\rm H})+{\cal O}([t_0-t_{\rm
H}]^2)$.  This provides us with the link between $U$ and $t_0$.%

In order to link $t_0$ with $u$, consider the integral on the right
hand side of Eq.~(\ref{uu}).  For $x\to +\infty$, the integrand
function vanishes, while near $\xi(t_0)$ it can be approximated by
$-c/(\kappa x)$.  Then the integral is just given by the difference
of the corresponding integrals evaluated at $x=+\infty$ and
$x=\xi(t_0)$, respectively, up to a possible finite positive
constant.  This gives $-\lambda\,(t_0-t_{\rm H})\sim
\mbox{const}\,\e^{-\kappa u}$.
Together with the previous link between $U$ and $t_0$ this leads to%
\begin{equation}%
U\sim U_{\rm H}-\mbox{const}\,\e^{-\kappa u}\;.%
\label{Uu'-bh}%
\end{equation}%
This relation between $U$ and $u$ is exactly the one found by
Hawking in his famous analysis of particle creation by a collapsing
star~\cite{hawking}. It is by now a standard result that this
relation implies the stationary creation of particles with a Planckian
spectrum at temperature $T_H=\kappa/(2\pi)$~\cite{bd,hu}.%

\noindent{\underline{\em Formation of a critical black hole}:}
Consider now the same type of function $\bar{v}(x)$ but this time
the sonic point is approached asymptotically in an infinite amount
of laboratory time.  Now the trapping horizon is just an asymptotic
point located at $x=0$, $t\to +\infty$, and in order to establish
whether an event horizon does, or does not, exist one must perform
an actual calculation of $U_{\rm H}$ for the ``last'' ray that
crosses the kink. The expression for $U_{\rm H}$ is again obtained
from Eq.~(\ref{UU}), noticing that now $t_0=+\infty$ along the
generator of the would-be horizon, so%
\begin{equation}%
U_{\rm H}=\lim_{t_0\to +\infty}\left(t_0
+\frac{1}{c}\int_{-\infty}^{t_0}\d
t\,\bar{v}\left(\xi(t)\right)\right)\;.%
\label{barU-cbh}%
\end{equation}%
The necessary and sufficient condition for the event horizon to
exist is that the limit on right hand side of Eq.~(\ref{barU-cbh})
be finite.  The integrand on right hand side of (\ref{barU-cbh}) can
be approximated, for $t\to t_0\to +\infty$, as $-c+\kappa\,\xi(t)$,
while for $t\to -\infty$ it just approaches zero.  Hence $U_{\rm H}$
is, up to a finite constant, equal to $\kappa/c$ times the integral
of $\xi$, evaluated at $t \to +\infty$.  For an exponential
behaviour, $U_{\rm H}$ turns out to be finite.%

For another right-moving sound ray that corresponds to a value
$U\lesssim U_{\rm H}$, combining Eqs.~(\ref{UU}) and
(\ref{barU-cbh}), realizing that $v\sim -c +\kappa \xi(t)$ for
$\xi(t)$ close to zero, and using expansion (\ref{appv-bh}), we find%
\begin{equation}%
U\sim U_{\rm
H}-\frac{\xi(t_0)}{c}-\frac{\kappa}{c}\int_{t_0}^{+\infty}\d
t\,\xi(t)\;.%
\label{deltaU'-cbh}%
\end{equation}%
For the asymptotically exponential $\xi(t)$ considered in this letter this implies
\begin{equation}%
U\sim
U_{\rm H}-\frac{A}{c}\left(1+\frac{\kappa}{\kappa_{\rm D}}\right)
\e^{-\kappa_{\rm D}\,t_0}.
\label{deltaUexp}%
\end{equation}%
For the link between $t_0$ and $u$ we obtain
\begin{equation}%
u\sim t_0-\frac{1}{\kappa}\,\ln\xi(t_0)\;,
\label{tu}%
\end{equation}%
as one can easily check by inserting the appropriate asymptotic
expansions into Eq.~(\ref{uu}).  Using Eqs.~(\ref{tu}) and
(\ref{deltaUexp}) we finally find%
\begin{equation}%
U\sim U_{\rm
H}-\mbox{const}\,\exp\left(-\frac{\kappa\,\kappa_{\rm D}}{\kappa
+\kappa_{\rm D}}\,u\right)\;.%
\label{Uuexp}%
\end{equation}%
Therefore, we have shown that by setting up a fluid flow that reaches
a sonic regime asymptotically with exponential rapidity in laboratory
time one can produce a stationary and Planckian spectrum of
quasi-particles at asymptotic infinity with temperature $T_{\rm eff} =
\kappa_{\rm eff}/(2\pi)$, where%
\begin{equation}%
\kappa_{\rm eff}:=\frac{\kappa\,\kappa_{\rm D}}{\kappa+\kappa_{\rm D}}\;.%
\label{kappaeff}%
\end{equation}%
When $\kappa_{\rm D}\gg \kappa$, $\kappa_{\rm eff}\simeq \kappa$ and
thus, our result is operationally indistinguishable from that of
Hawking.  In this sense neither the existence of an ergoregion nor
that of an apparent horizon are needed for the simulation of Hawking
radiation.  Remarkably this is true also for a ``double-sided''
velocity profile in which the asymptotic sonic regime is limited to a
single spatial point~\cite{quasi-particle-cqg}.%

In terms of inverse temperatures, defined as 
$\beta_i = 1/T_i = 2\pi /\kappa_i$, we have the very suggestive result 
\begin{equation}%
\beta_{\rm eff} = \beta_{\rm Hawking} + \beta_{\rm D}.
\end{equation}%

\noindent{\underline{\em Experimental realizability}:} The critical
black hole model seems worth taking into consideration in connection
with the realizability of a Hawking-like flux in the laboratory. The
creation of supersonic configurations in a laboratory is usually
associated with the development of instabilities.  There are many
examples of the latter in the literature; \eg\ in
Ref.~\cite{volovik-ripplons} it was shown that in an analogue
spacetime based on ripplons on the interface between two different
sliding superfluids (for instance, $^3$He-phase A and $^3$He-phase
B), the formation of an ergoregion would make the ripplons acquire
an amplification factor that eventually would destroy the
configuration. Therefore, this analogue system, although very
interesting in its own right, will prove to be useless in terms of
detecting a Hawking-like flux.  However, by creating, instead of an
ergoregion, a critical configuration one should be able to at least
have a better control of the incipient instability, while at the
same time producing a dynamically controllable Hawking-like flux.

Nevertheless, the actual realization of a critical configuration might
also appear problematic for entirely different reasons. The
corresponding velocity profiles are characterized by discontinuities
in the derivatives, so one might wonder whether they would be amenable
to experimental construction, given that the continuum model is only
an approximation.  In particular, could it be that in a real
experimental setting one does not observe the results predicted by our
analysis, but those corresponding to models where the discontinuities
have somehow been smoothed out?%

Fortunately, this will not be the case. In realistic situations,
what is relevant for Hawking radiation is a coarse-grained profile
--- obtained by averaging over a scale $\Delta$ larger than the one
which characterizes the breakdown of the continuum model --- which
does not contain the unphysical small scale details of $v(x,t)$.
This implies that the reliable results are those involving features
recognizable over length scales of order $\Delta$, or
larger~\cite{quasi-particle-cqg}.  In particular, the relevant
surface gravity will be defined by averaging the slope of the
velocity profile over scales which are of order of $\Delta$.  This
averaged surface gravity will be non-zero for the critical profile,
as well as for a smoothed one.  Indeed, it will be approximately
equal to the surface gravity at the horizon of the critical black
hole, while it will obviously not coincide with the one of the
smoothed profile (which is zero)~\cite{quasi-particle-cqg,footnote}.%

\noindent{\underline{\em Hints for semiclassical gravity}:}
The critical collapse result also suggests an alternative scenario for
the semiclassical collapse and evaporation of black hole-like objects.
Although somewhat speculative at this stage we believe it is
worthwhile to describe it here.%

Imagine a dynamically collapsing star.  The collapse process starts
to create particles dynamically before the surface of the star
crosses its Schwarzschild radius. (This particle creation is
normally associated with a transient regime and have nothing to do
with Hawking's Planckian radiation.) The energy extracted from the
star in this way will (due to energy conservation) reduce its total
mass, and so also its Schwarzschild radius, so $\kappa$ increases.
By this argument alone, we can see that a process is established in
which the surface of the star starts to ``chase'' its Schwarzschild
radius while both collapse towards zero (this situation was already
described by Boulware in Ref.~\cite{boulware}).%

Now, contrary to the standard view, it could happen that the
back-reaction on the geometry is such that it prevents the surface
of the star from actually crossing its Schwarzschild radius
(see~\cite{stephens-thooft} for a discussion of this possibility).
The function $\xi(t)$ in our calculation would represent the radial
distance between the surface of the star and its Schwarzschild
radius. Then it is sensible to think that during the evaporation
process any dynamical $\kappa_{\rm D}$ would also depend on $t$. As
the evaporation temperature increases ($\kappa$ increases) the
back-reaction would become more efficient and therefore we might
expect that $\kappa_{\rm D}$ decreases. Then the evolution of the
evaporation temperature would interpolate between an early-time
temperature completely controlled by $\kappa(t)$, and a late-time
temperature completely controlled by $\kappa_{\rm D}(t)$. Therefore,
this black hole-like object will have a temperature-decreasing
phase, showing a possible semiclassical mechanism for regularizing
the end point of the evaporation process.%

In this scenario the complete semiclassical geometry will have
neither a trapped region (and so by construction no trapping
horizon) nor an event horizon. In this circumstance there would be
no trans-Planckian problem, nor information loss associated with the
collapse and evaporation of this black hole-like object. Whether
this scenario is viable or not will be the subject of future work.



\begin{thebibliography}{99}
\bibitem{footnote}%
Of course, in addition, in any realistic system one will have to
take into consideration the existence of modified dispersion
relations. This does not seem to alter the above
scenario~\cite{unruh}. (For a discussion of these issues we refer
the reader to~\cite{quasi-particle-cqg}.)%
\end{thebibliography}
\end{document}